# Large electro-opto-mechanical coupling in VO$_2$ neuristors

Upanya Khandelwal[1], Rama Satya Sandilya[1], Rajeev Kumar Rai[1,2], Deepak Sharma[1], Smruti Rekha Mahapatra[3], Debasish Mondal[3], Navakanta Bhat[1], Naga Phani Aetkuri[3], Sushobhan Avasthi[1], Saurabh Chandorkar[1*], Pavan Nukala[1*]

**Affiliations:**

[1] Center for Nanoscience and Engineering, Indian Institute of Science; Bengaluru, Karnataka, India, 560012.

[2] Materials Science and Engineering, University of Pennsylvania; 3231 Walnut Street, Philadelphia, 19104.

[3] Solid State and Structural Chemistry Unit, Indian Institute of Science; Bengaluru, Karnataka, India, 560012.

*Corresponding authors: Pavan Nukala, email: pnukala@iisc.ac.in

Saurabh Chandorkar, email: saurabhc@iisc.ac.in

**Abstract:** Biological neurons are electro-mechanical systems, where the generation and propagation of an action potential is coupled to generation and transmission of an acoustic wave. Neuristors, such as VO$_2$, characterized by insulator-metal transition (IMT) and negative differential resistance, can be engineered as self-oscillators, which are good approximations of biological neurons in the domain of electrical signals. In this study, we show that these self-oscillators are coupled electro-opto-mechanical systems, with better energy conversion coefficients than the conventional electromechanical or electrooptical materials. This is due to the significant contrast in the material's resistance, optical refractive index and density across the induced temperature range in a Joule heating driven IMT. We carried out laser interferometry to measure the opto-mechanical response while simultaneously driving the devices electrically into self-oscillations of different kinds. We analyzed films of various thicknesses, engineered device geometry and performed analytical modelling to decouple the effects of refractive index change vis-à-vis mechanical strain in the interferometry signal. We show that the effective piezoelectric coefficient ($d_{13}^*$) for our neuristor devices is 660$\pm$20 pm/V, making them viable alternatives to Pb-based piezoelectrics for MEMS applications. Furthermore, we show that the effective electro-optic coefficient ($r_{13}^*$) is ~22 nm/V, which is much larger than that in thin-film and bulk Pockels materials.

**One sentence summary:** VO$_2$ neuristors are coupled electro-opto-mechanical systems, with better energy conversion coefficients than the conventional electromechanical or electro-optical materials.

# INTRODUCTION

Action potentials (AP) in biological neurons, measured through electrophysiological techniques, are electrical signals that emerge and propagate along the axonal membrane(*1*, *2*). Significant evidence suggests that the AP generation and propagation is accompanied by rapid and transient mechanical alterations(*3*). These result in mechanical changes including modifications in the axonal radius(*4*, *5*), the release and reabsorption of a small amount of heat(*6*), pressure(*7*) and contraction of the axon at its terminus upon the arrival of the AP(*8*). It has been shown that close to the temperatures of physiological interest, a chain melting phase transition in the neuronal membrane accompanied by a change in density and heat capacity, is responsible for the creation and propagation of mechanical displacements and sound waves, which cause the aforementioned alterations(*9*, *10*). It has been argued that the change in membrane capacitance during such a phase transition can explain the associated action potential (contrasting hypothesis to Hodgkin-Huxley models)(*11*), making neurons coupled electro-mechanical systems(*12*).

Correlated electronic systems(*13*) such as $VO_2$ exhibit temperature dependent first-order phase transitions, analogous to lipid membranes(*14*). This is an insulator-to-metal transition (IMT), accompanied by a structural change from monoclinic to rutile phase(*15*), signified by substantial contrast in electrical resistance(*16*, *17*), density(*18*) and refractive index(*19*, *20*) between the two phases. In a device setting, this transition occurs through current-induced Joule heating(*21*), resulting in volatile switching in voltage-controlled I-V measurements, and a negative differential resistance (NDR) region in current-controlled measurements(*22*). It has been shown by several groups, that by suitably operating the devices in the NDR regions, self-oscillations can be engineered (*23–26*). Infact, Yi et al., (*27*) have shown 23 different types of electrical signals in this neuristor system, emulating a complex portfolio of electrical activity exhibited by biological neurons. In particular, three types of neuronal signals i.e. tonic spiking, tonic bursting and phasic spiking are of interest to the current work. Tonic spiking refers to a sustained oscillatory response, throughout the period of stimulus, and is observed in retinal ganglion cells(*28*) and vertebrate olfactory epithelium(*29*). Tonic bursting refers to clusters of tonic spiking interspersed with periods of non-activity (resting) during a single stimulus. Interneurons(*30*) and thalamic neurons(*31*) are examples where tonic bursting is generally observed. Phasic spiking refers to a transient response with one or few action potentials at the outbreak of stimulus followed by resting, and is also seen in retinal ganglion neuron cells (*28*).

In $VO_2$ neuristors, thus far the analogy with biological neurons has been shown only in the electrical domain(*27*, *32*). This motivated us to ask the following questions: a) whether $VO_2$ neuristors are also coupled electro-mechanical systems, especially given the large volume change occurring during IMT, b) whether these are coupled electro-optical systems, given the refractive index changes across IMT, and c) if so, more importantly, what is the magnitude of energy conversion coefficients in both the domains, and how would they compare with conventional piezoelectric and electrooptic materials. To that end, we drove the devices electrically using a DC bias into self-oscillations, and simultaneously measured the opto-mechanical response using laser interferometry. We show tonic spiking, tonic bursting and phasic spiking behaviors (from 100s kHz to MHz frequencies), and clearly demonstrate a coupled opto-mechanical response with the same power spectrum as that of the electrical oscillations, and significantly large coupling coefficients. Through systematic experiments based on varying sample thickness, device designs and analytical modelling, we understand the contributions of mechanical and optical effects to the combined response.

**RESULTS AND DISCUSSION**

Thin films of $VO_2$ were grown epitaxially on sapphire (and $TiO_2$) substrates via Pulsed Laser Deposition (PLD) (see supplementary methods for details of the growth). Cross-sectional electron transparent samples were prepared using focused ion beam technique. STEM-EELS depth profiling analysis was carried out on these lamellae to understand the chemical inhomogeneities in the film (details in Supplementary text S1, Fig. S1). $L_{2,3}$ of peaks of Vanadium appear at lower energies in about the first ~10 nm close to the surface of the film, compared to the bulk, consistent with the surface layer being at a lower oxidation state than bulk(*33*) .This interpretation is further in line with the larger $L_3/L_2$ (white line ratios) at the surface than the bulk. (Fig. S1).

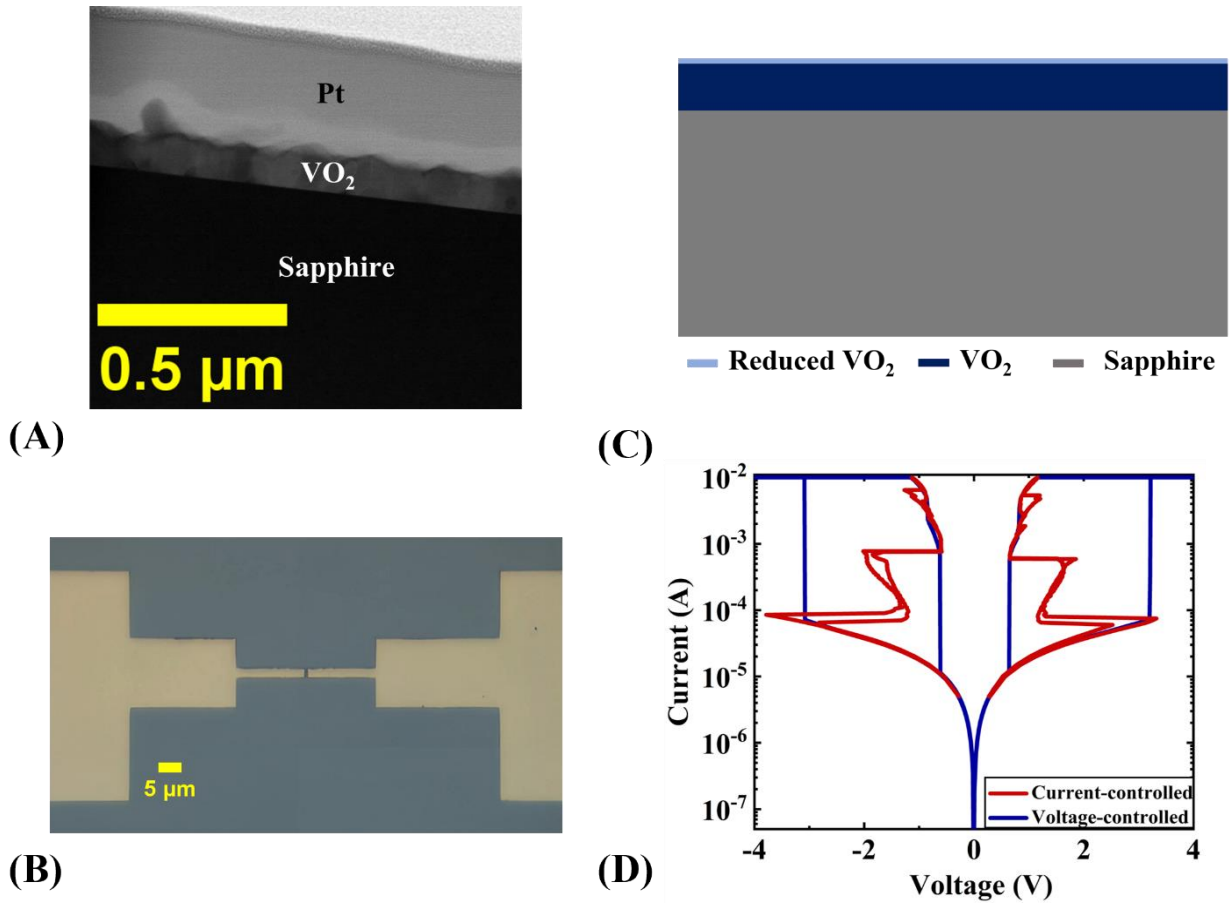

**Fig. 1**: **Correlating VO₂ chemistry with I-V characteristics** (**A**) Cross-sectional TEM image of $VO_2$ on Sapphire. (**B**) Optical micrograph of the fabricated device. (**C**) Schematic illustrating that $VO_2$ film on Sapphire is a heterostructure consisting of a surface layer (~10 nm) with V in a slightly lower oxidation state, and bulk layer (for EELS and transport data that support this conclusion see Fig. S1 and Supplementary text S1). (**D**) Current-Voltage characteristics of the device, in both current controlled (red) and voltage controlled (blue) modes.

On this platform, we fabricated two terminal lateral nano-gap devices (Fig. 1(B)) using e-beam lithography (gap sizes: 500 nm to 900 nm), with Pt electrodes (see supplementary methods) similar to the devices discussed in (*34*). We measured resistance as a function of temperature (R-T) on representative devices and found atleast two different transition temperatures (Fig. S2), both close to the bulk $T_{IMT}$ in $VO_2$ (68°C). Our R-T measurements along with the EELS depth profiling, allow us to propose a simple model to visualize our thin-film as a heterostructure of slightly reduced $VO_2$ (~10 nm)/$VO_2$ (~90 nm) on Sapphire substrate (Fig. 1(C)).

The voltage-controlled I-V measurements show a volatile switching, and current-controlled I-V profiles exhibit multiple negative differential resistance (NDR) regions (Fig. 1(D)). We simulated the double NDR behavior by slightly adapting the electro-thermal model (ETM) developed by us in reference (*35*) (also see Fig. S3), by including two heat sources, one corresponding to $VO_2$ layer, and another to the reduced $VO_2$ layer. The details of this simulation are described in the Supplementary text S2 and S3. More generally, our modelling allows us to suggest that multiple NDRs observed in various neuristors (*32, 36*) may be a result of heterogeneity in the cationic oxidation state in these films.

To observe the current oscillations, a 20 kΩ biasing resistor was connected in series with the device (along with parasitic capacitance) and voltage pulses were applied (see supplementary methods). In Fig. S4 we show electrical oscillations on a representative device (gap size: 700 nm) among all the tested devices (~20), as a function of applied voltage ($V_a$). At $V_a < 9$ V, we see a simple capacitive charging and discharging of the $VO_2$ neuristor (Fig. S4(A)). However, for 10 V < $V_a$ < 13 V, we observe current oscillations in the form of tonic spiking (Fig. S4(B)), which exhibits frequency encoding behavior of voltage, typical of neuristors (Fig. S5)(*35*). These electrical oscillations are manifestations of device temperature oscillations across the phase transition between insulating and metallic phases in Mott insulators, and were simulated using our ETM framework ((*35*), Fig. S3 (D)). Interestingly, for 13.1 V < $V_a$ < 13.8 V, bunched activity of current oscillations, also known as tonic bursting behavior (Fig. S4(C)) appears. Finally, at $V_a$> 13.8 V, phasic spiking (or initial spiking followed by non-activity) is observed (Fig. S4(D)). Thus, in a single neuristor (not coupled to any other neuristor as in (*27*)) we show three different neural spiking behaviors that evolve with the application of external voltage (tonic spiking to tonic bursting to phasic spiking). Tonic bunching is captured in ETM upon inclusion of floating terminals, modeled as parallel inductors and capacitors, as shown in Figs. S6(A and B) (also see supplementary text S3). It may be noted that Yi *et al*., have shown 23 different spiking behaviors using two coupled $VO_2$ neuristors (*27*).

The insulator to metal transition (IMT) is also signified by a corresponding structural phase transition between the monoclinic phase and the rutile phase of $VO_2$, which involves volumetric strain (mechanical displacement) and refractive index contrast. To understand and measure the possible coupled electro-mechanical and electro-optical nature of these signals, we synchronized our laser interferometry set up (in a Laser Doppler Vibrometer (LDV) system) with the parameter analyzer, as shown in Fig. S7 (also see supplementary methods). We simultaneously measured the current oscillations through the parameter analyzer and optical path difference (OPD) oscillations (originating from mechanical displacements and refractive index oscillations, to be referred to as effective displacement from here on) through the interferometer as shown in Fig. 2(A) (see supplementary text S4). For a representative device (D1, with channel length 850nm) set to tonic spiking in electrical domain (Fig. 2(B)), corresponding (filtered) opto-mechanical response of devices is shown in Fig. 2(C) [see supplementary text S4, Figs. S9(A-D) for raw signal and filtering procedures], with the amplitudes of effective displacement oscillation ~5 nm. The fundamental frequency and the harmonics of the opto-mechanical oscillations (Fig. 2(E)) are strikingly similar to the electrical oscillations (Fig. 2(D)), corroborating their coupled electro-opto-mechanical nature. For another representative device D2, with channel length 700 nm set to tonic bursting, the temporal profiles of both the current oscillations (Fig. 2(F)) and the effective displacement oscillations (Fig. 2(G)) are the same.

Next, we designed structures and performed experiments to uniquely understand and isolate only the coupled electro-mechanical displacements (from electro-optical effect). The device design is shown schematically in Fig. S10(A), where the device is first coated by an insulator followed by a metal on the top (supplementary methods). The metal does not allow light to penetrate into the $VO_2$ layer, and the insulator isolates the device electrically from the top metal. Thus, any measured path differences in the interferometry signal can only be attributed to the device mechanical displacements. This structure was achieved by spin coating PMMA (polymethyl methacrylate) as an insulating layer, and patterning Pt metal on the top using focussed ion beam technique (see supplementary methods and Fig. S10).

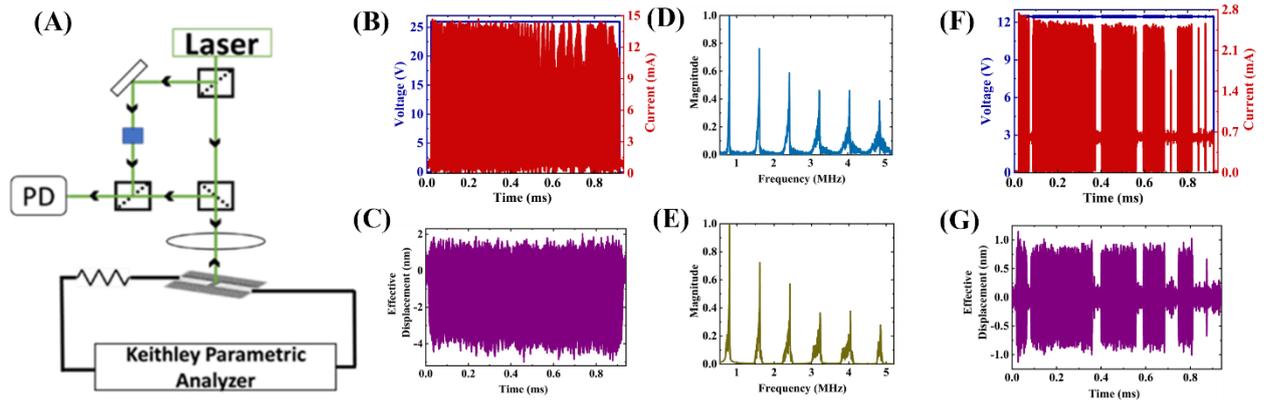

**Fig. 2**: **Electrical and Opto-Mechanical oscillations.** (**A**) Schematic of measurement setup for synchronized electrical and opto-mechanical measurements. (**B**) Tonic spiking of current oscillations of device D1 with channel length 850nm. (**C**) Filtered tonic spiking of optical path difference (effective displacement) signal measured using LDV (see supplementary text S4 for details on filtering and raw signals). (**D**) Normalized FFT of current oscillations shown in (B). (**E**) Normalized FFT of filtered effective displacement shown in (C). (**F**) Tonic Bursting of current oscillations. (**G**) Tonic Bursting of displacement oscillations (raw signal) on a representative device D2 with channel length 700nm.

Upon geometrically engineering the same representative device discussed in Fig. 2, and resetting it to tonic spiking (Fig. 3(A), see FFT in Fig. 3(C)), the effective displacement amplitude (noise filtered) which corresponds solely to mechanical displacement reduced to ~0.22 nm (Fig. 3(B), see FFT in Fig. 3(D), also see supplementary text S4 for details on filtering and raw signals, Fig. S9 (E-H)).

To validate if such mechanical displacements are reasonable, we performed θ-2θ x-ray diffraction measurements, while heating our $VO_2$ thin-films *in situ* across the phase transition temperature ~68°C (Fig. 3(E)). The 2θ decreases from 39.89° at 30°C (monoclinic phase) to 39.80° at 100°C (rutile phase). The complete evolution of 2θ as a function of temperature is shown in Fig. S11. This corresponds to 0.22% change in the out-of-plane lattice parameter, or 0.22 nm displacement amplitude for a 100 nm thick film, similar to our interferometry measurements (Figs 3(B) and S9 (D)).

So far, we reported the current oscillations across the device, when it is set to self-oscillations. However, to compare the electro-mechanical effect with standard piezoelectric materials, we independently measured the voltage oscillations across several DUTs of various channel lengths. Fig S12 shows measured voltage oscillations on a DUT of ~280 nm channel length (see also supplementary text S5) corresponding to a current oscillation of ~7 mA and mechanical displacement of ~0.22 nm. Using this, we estimated an effective piezoelectric coefficient ($d_{13}$*) in all these devices to be ~660 ±20 pm/V (see supplementary text S4 and Fig. S12 (D) for estimation and error analysis), which is larger than piezoelectric thin films such as PZT (~150 pm/V)(*37*). Thus, more fundamentally, our results suggest that large electromechanical effects with responses larger than Pb-based piezoelectric materials can be engineered with phase-changing neuristors such as $VO_2$. It is worth noting that both the metallic (rutile) and the insulating (monoclinic) phases of $VO_2$ are both centrosymmetric.

To uniquely quantify the refractive index oscillations experimentally, and assess these materials for electro-optic applications, we performed synchronized interferometry on another set of devices fabricated on ultra-thin $VO_2$ films (~10 nm), epitaxially grown on $TiO_2$ (101) ((*38*), see methods). Owing to their ultra-thin nature, the mechanical displacements on these samples are negligible. Fig. 4 shows data on a representative device, with filtered effective displacement oscillation amplitude of ~2.5 nm (also see Figs. S9 (I-L)), corresponding to a refractive index contrast of $\Delta n$= 0.115, between the metallic and insulating phases in this system (as obtained from analytical modelling framework described next). Electro-optic(EO) coefficient defined as $r_{13}= \Delta n/E_{max}$, where $E_{max}$ is the amplitude of electric field oscillation in the DUT (see measured voltage oscillation data in Fig. S14 corresponding to a current oscillation of ~5.1 mA) is estimated as ~22 nm/V. Note that largest Pockels coefficients were reported on epitaxial thin films of $BaTiO_3$, and these values $r_{42}$~943 pm/V(*39*), are much smaller in comparison with EO coefficients on phase-change neuristors.

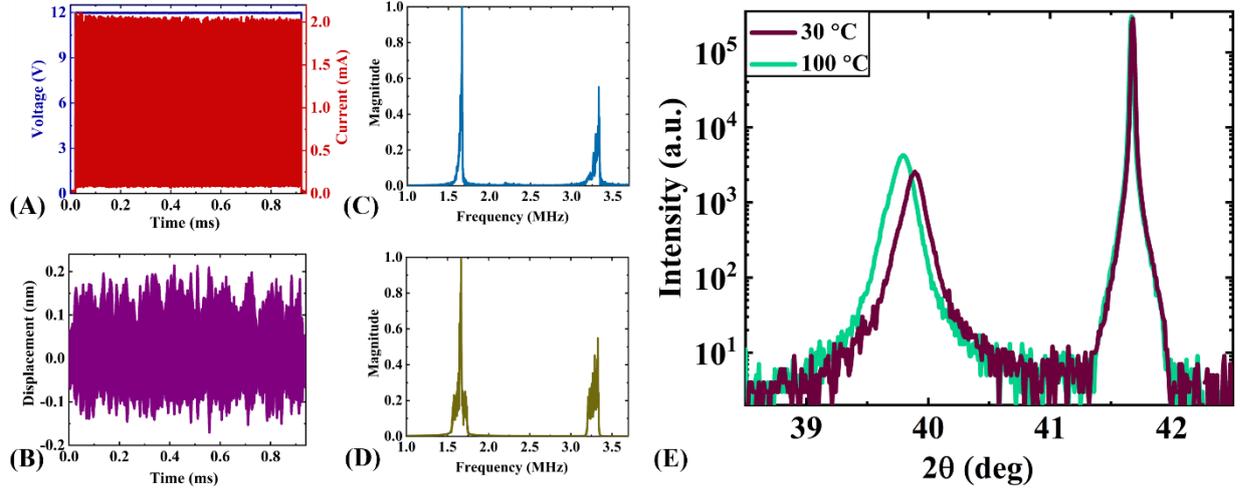

**Fig. 3**: **Isolating mechanical oscillations.** (**A**) Current oscillations after depositing metal on top of PMMA. (**B**) Filtered displacement oscillations corresponding to (A) (see supplementary text S4 for raw signal and filtering procedure). (**C**) Normalized FFT of current oscillations in (A). (**D**) Normalized FFT of effective filtered displacement oscillations in (B). (**E**) In-situ XRD at 30°C and 100°C.

Next, we analytically modelled the path difference oscillations using optical transfer matrix framework to gain insights into the individual contributions of refractive index changes vs mechanical displacements (supplementary text S4). Given that our phase transition oscillations are non-linear (with Fourier components at $\Omega, 2\Omega, 3\Omega,...$, see FFT of current oscillations in Fig. 2(D)), we considered the change in refractive index ($N+iK$) upon phase transition from insulator-to-metal upto 4$^{th}$ harmonic as follows:

$$N = \frac{n_1+n_2}{2} + \frac{n_1-n_2}{2}(A_1 \sin(\Omega t) + A_2 \sin(2\Omega t) + A_3 \sin(3\Omega t) + A_4 \sin(4\Omega t))$$
Eq.1

$$K = \frac{k_1+k_2}{2} + \frac{k_1-k_2}{2}(A_1 \sin(\Omega t) + A_2 \sin(2\Omega t) + A_3 \sin(3\Omega t) + A_4 \sin(4\Omega t))$$
Eq.2

where, $n_1$ and $n_2$ are the refractive index in insulating phase and metallic phase respectively(*20*), $A_1, A_2, A_3$ and $A_4$ are the Fourier components obtained from experimental current oscillations of a representative device (Fig. 2(C)), and $\Omega$ is the frequency of current oscillations.

The mechanical oscillation of the membrane is also captured by the same Fourier components as follows:

$$\Delta s = d_s * (A_1 \sin(\Omega t) + A_2 \sin(2\Omega t) + A_3 \sin(3\Omega t) + A_4 \sin(4\Omega t)) \quad \text{Eq.3}$$

where $d_s$ is chosen to be 0.11 nm (based on in-situ XRD data in Fig. 3(E)).

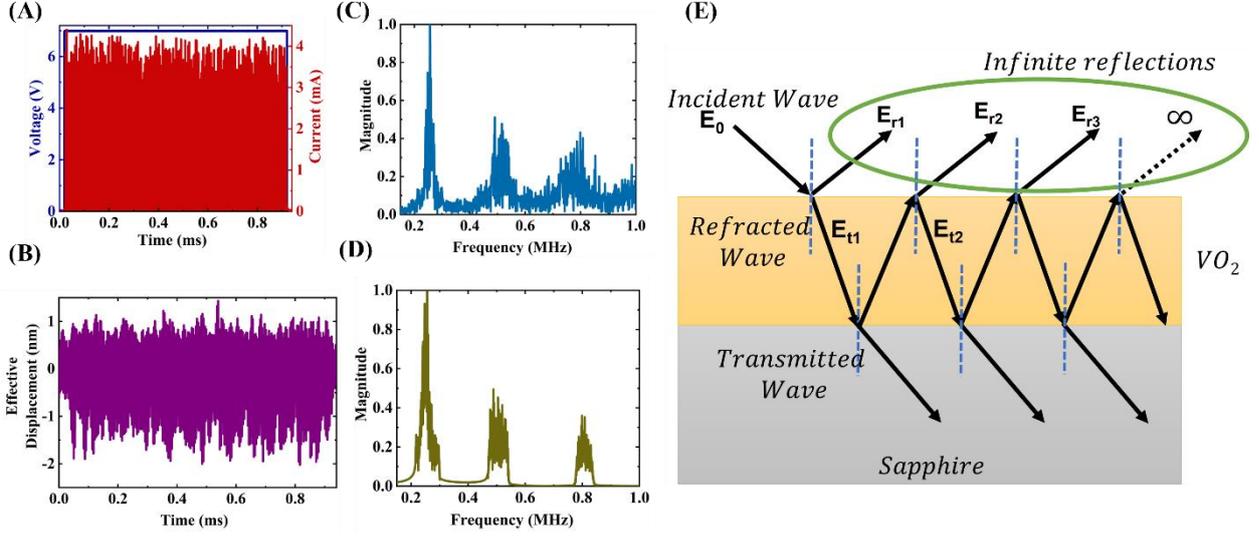

**Fig. 4**: **Probing refractive index oscillations on ultrathin films.** (**A**) Current oscillations of $VO_2$ (10 nm) on $TiO_2$. (**B**) Filtered effective displacement oscillations corresponding to current oscillations in (A). (**C**) FFT of current oscillations in (A). (**D**) FFT of effective displacement oscillations in (B). (**E**) Schematic of multiple reflections from thin film.

Assuming normal incidence with $VO_2$ film on Sapphire (Fig. 4(E)) we solved this problem using optical transfer matrix formalism. When light with frequency ω strikes ($E_0 = E_{00} \sin(\omega t)$) on a surface of thin film ($VO_2$), it either gets reflected or transmitted and absorbed (Reflected component: $E_{r1}$, Transmitted component: $E_{t1}$). The membrane mechanical oscillation is solely captured in first reflection ($E_{r1}$) as follows:

$$E_{r1} = \frac{n_{VO_2} - n_{air}}{n_{VO_2} + n_{air}} \, e^{i(\frac{2\pi}{\lambda}\Delta s * \sin(\Omega t))} E_0$$
Eq.4

After multiple reflections(r) and transmissions(t) (i=2, 3, ...inf), electric field components from the film interface can be written as,

$$E_{t1} = \frac{\sqrt{4 n_{air} n_{VO_2}}}{n_{VO_2} + n_{air}} \, e^{i(\frac{2\pi}{\lambda}(N+iK)s + \Delta s * \sin(\Omega t))} E_0$$
Eq.5

$$E_{r2} = \frac{\sqrt{4 n_{air} n_{VO_2}}}{n_{VO_2} + n_{air}} \frac{n_{Sapphire} - n_{VO_2}}{n_{Sapphire} + n_{VO_2}} \, e^{i(\frac{2\pi}{\lambda}(N+iK)s)} E_{t1}$$
Eq.6

$$E_{t2} = \frac{n_{VO_2} - n_{air}}{n_{VO_2} + n_{air}} \frac{n_{Sapphire} - n_{VO_2}}{n_{Sapphire} + n_{VO_2}} \, e^{i(\frac{4\pi}{\lambda}(N+iK)s)} e^{i(\frac{2\pi}{\lambda}\Delta s * \sin(\Omega t))} E_{t1}$$
Eq.7

$$E_{r3} = \frac{\sqrt{4n_{air}n_{VO_2}}}{n_{VO_2}+n_{air}} \left(\frac{n_{Sapphire}-n_{VO_2}}{n_{Sapphire}+n_{VO_2}}\right)^2 \frac{n_{VO_2}-n_{air}}{n_{VO_2}+n_{air}} e^{i\left(\frac{6\pi}{\lambda}(N+iK)s+\frac{4\pi}{\lambda}\Delta s*\sin(\Omega t)\right)} E_{t1}$$

Eq.8

After infinite reflections at the film-air interface, and film-substrate interface, the signal E∞ (reaching the photodetector) is given by Eq.9

$$E_\infty = E_{r1} + \frac{E_{r2}}{1-\frac{E_{r3}}{E_{r2}}}$$

Eq.9

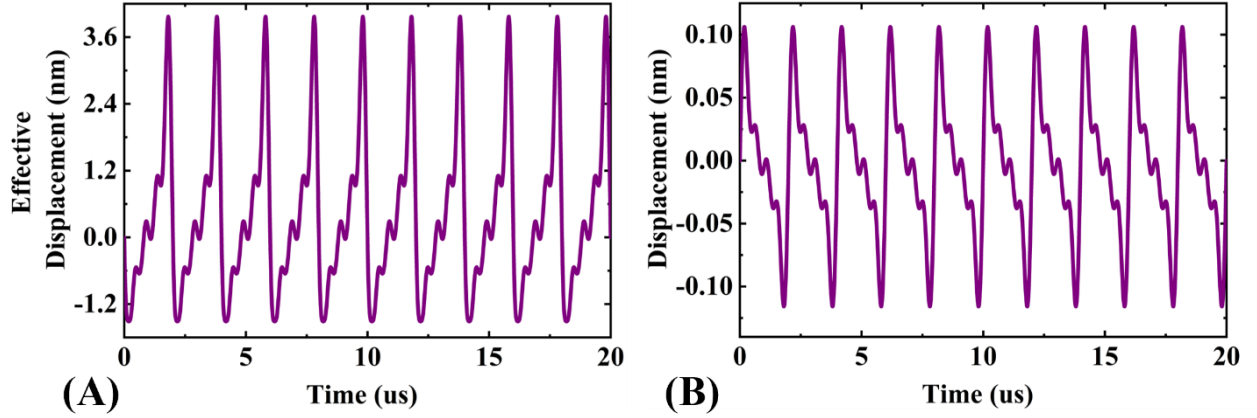

Fig. 5: **Analytical simulations for refractive index oscillations on 100 nm thick films** (**A**) Effective displacement oscillations including the effects of refractive index and mechanical oscillations. (**B**) Mechanical displacement considering only $E_{r1}$ of $VO_2$ (100 nm) on sapphire.

We calculate that the effective displacement oscillations (related to the phase of E∞ vs time shown in Fig. 5(A), also see Supplementary text S4) have an amplitude of 5.46 nm (similar with our measurements, Fig. 2(B)). The amplitude of mechanical oscillations is 0.22 nm ($E_{r1}$ vs time in Fig. 5(B)) in accordance to our measured displacements (Fig. 3(B)). Analytical simulations were also carried out for ultra-thin film $VO_2$(10 nm)/$TiO_2$ (shown in Fig. S15). These results also issue a cautionary note on the interpretations of laser interferometry measurements on transparent electromechanical samples, in which electro-optic effects need to be considered. Conversely, our simple modelling framework, combined with synchronized interferometry, can be a simple and useful tool to estimate electro-optic refractive index contrast in newer materials systems.

## CONCLUSION

In this work, we show that $VO_2$ neuristors are not only coupled electro-mechanical systems similar to biological neurons but are coupled electro-opto-mechanical systems. First, we show in electrical domain, three different types of neuronal oscillatory behaviors from a single $VO_2$ neuristor device. By adapting an electrothermal modelling framework developed earlier by us(35),and taking insights from our structural data about the chemistry of $VO_2$, we simulated the various oscillatory behaviors engineered in our $VO_2$ neuristors. By synchronizing our electrical system with laser interferometry measurements, we show corresponding opto-mechanical response of $VO_2$ when driven into all these types of oscillations. By a suitable geometric design of the films, we were able to measure mechanical displacement amplitudes of these self-oscillators of ~0.22 nm, with their electro-mechanical response (~660 pm/V) larger than or comparable to piezoelectric Pb-based materials. In ultrathin films, we established a simple way of estimating refractive index contrast across the phase transition, and estimated the electro-optic coefficients to be ~25 nm/V much larger than conventional Pockels materials. Finally, we analytically modelled simple lateral two terminal devices to separate the contribution of optical response from mechanical response in one device, using optical transfer matrix formalism, and understood that refractive index modulations in these neuristors are significant. That neuristors are important neuromorphic elements has been already well established. The magnitude of effective electromechanical and electrooptic coefficients we were able to sensitively measure opens up $VO_2$ neuristors as promising devices for MEMS and electrooptic applications. We advertise for $VO_2$ as a great candidate for neuromorphic computing in electro-opto-mechanical domains, using MEMS and electrooptic platforms.

**Acknowledgements:** This work was partly carried out at Micro and Nano Characterization Facility (MNCF), and National Nanofabrication Center (NNfC) located at CeNSE, IISc Bengaluru, funded by NPMAS-DRDO and MCIT, MeitY, Government of India; and benefitted from all the help and support from the staff. P.N. acknowledges Start-up grant from IISc, Infosys Young Researcher award, and DST-starting research grant SRG/2021/000285. The authors acknowledge funding support from the Ministry of Human Resource Development (MHRD) through NIEIN project, from Ministry of Electronics and Information Technology (MeitY) and Department of Science and Technology (DST) through NNetRA and the Thematic Unit of Excellence for Nano Science and Technology project from DST Nano Mission. All the authors acknowledge the usage of national nanofabrication center, micro nano characterization center, and advanced facility for microscopy and microanalysis of IISc for various fabrication and characterization studies.

# Supplementary Materials for

## Large electro-opto-mechanical coupling in VO$_2$ neuristors

Upanya Khandelwal[1], Rama Satya Sandilya[1], Rajeev Kumar Rai[1,2], Deepak Sharma[1], Smruti Rekha Mahapatra[3], Debasish Mondal[3], Navakanta Bhat[1], Naga Phani Aetkuri[3], Sushobhan Avasthi[1], Saurabh Chandorkar[1*], Pavan Nukala[1*]

*Corresponding authors:  Pavan Nukala, email: pnukala@iisc.ac.in
Saurabh Chandorkar, email: saurabhc@iisc.ac.in

**The PDF file includes:**

Materials and Methods
Supplementary Text
Figs. S1 to S15
Table S1 to S2
References

**Materials and Methods**

**Synthesis by Pulsed Laser Deposition**

Growth of $VO_2$ on Sapphire:

$VO_2$ thin films were grown on $Al_2O_3$(0001) substrates through pulsed laser deposition (PLD). $V_2O_5$ powders (99.6% purity) were pressed at a load of 8 tons into a shape of 1 inch diameter pellet which is then sintered at 600ºC in ambient atmosphere for 12hrs. This sintered $V_2O_5$ pellet is used as a PLD target. $Al_2O_3$ (sapphire) substrates were cleaned thoroughly by standard solvent cleaning process (acetone, IPA and DI water) before loading into the PLD chamber. The chamber is pumped down to a base pressure of 5e-6 mbar. Substrate is then heated to 700˚C. The deposition pressure of 3e-2 mbar was maintained by flowing $O_2$ gas into the chamber. $V_2O_5$ target was ablated by KrF excimer laser with an energy fluence close to 1J/cm2. Laser repetition rate, duration of the pulse, spot size and the energy of the laser pulse on the target are 3 Hz, 25 ns, 9 mm$^2$ and 90 mJ respectively.

Growth of $VO_2$ on $TiO_2$:

$VO_2$ thin films were deposited on single-crystalline rutile $TiO_2$ (101) substrates (Shinkosha, Japan) using PLD (NEOCERA) with a 248 nm KrF laser. Prior to deposition, single crystalline $TiO_2$ (101) substrates were treated using the procedure reported previously(*1*). A sintered $V_2O_5$ pellet was used as the target. The chamber was pumped down to a base pressure of ~5×10$^{-8}$ Torr. The laser energy density during deposition was ~ 1.5 J/cm$^2$, a laser repetition rate of 2 Hz, deposition oxygen pressure of 10 mTorr and the substrate temperature of 425 °C was used for all the depositions. Under these conditions, the growth rate was calculated to be ~5×10-2 Å/pulse.

**TEM imaging**

Electron transparent cross-sectional samples were prepared using focused ion beam (FIB). High-angle annular dark-field (HAADF) scanning transmission electron microscopy (STEM) imaging and electron energy loss spectroscopy (EELS) were performed using a probe corrected TITAN Themis microscope (60-300 kV) operating at 300 kV with a convergence angle of 24.5 mrad. HAADF-STEM images were acquired at 160 mm camera length with a probe current of 160 pA. For EELS, a probe current of 300 pA, 38 mm camera length and 0.1 eV/channel dispersion were used to maximize the signal-to-noise ratio and sufficient energy resolution. A GIF quantum ER camera provided by GATAN Inc. was used to collect the EELS spectra, using a 5 mm collection aperture on Gatan imaging filter (GIF). The images and spectra were further processed with Gatan Digital Micrograph software. Standard EELS tuning procedure was employed before acquiring real spectra, with the width of the Zero Loss Peak being 0.8 eV.

**Device fabrication**

Two terminal lateral devices were fabricated on $VO_2$/Sapphire and $VO_2$/$TiO_2$ using e-beam lithography (EBL). Samples were first spin-coated with PMMA 495A4 followed by patterning for electrodes via EBL with channel lengths ranging from 700-900 nm. After developing the patterned electrodes, 30nm Platinum (Pt) was sputtered for the metal contacts followed by lift-off. The optical image of the patterned device is shown in Fig. 1(B).

**Experimental setup for measuring current oscillations**

To measure current oscillations, we use Keithley 4200 SCS parameter analyzer (PA) with two PMU's (Pulse measuring units). PMU 1(used to generate voltage pulses) was connected to a bias resistor of 20kΩ and this bias resistor is then connected to one terminal of the device. The other terminal of the device was connected to PMU 2 which was grounded. Current was measured through PMU 2.

**Synchronized setup for Current and Displacement Measurements**

For measuring current oscillations, LDV and PA were synchronized to measure current and displacement simultaneously (Fig. 2A). To synchronize LDV with PA, trigger output of the PA was fed to the reference channel of LDV setup. Optical path difference oscillations were measured as an output of displacement from LDV. Measurement setup is shown in Fig. S7.

**Device fabrication to measure the effect of only mechanical motion**

The device fabricated as described in the device fabrication section was used and coated with 80nm thick e-beam resist PMMA 950A2. Patterning was done using EBL to develop the regions for probing the electrodes. After this 50-80 nm e-beam Pt was deposited using focused ion beam (FIB). Schematic and SEM image shown in Fig. S10 (A and B) respectively.

**Supplementary Text**

**S1. STEM-EELS Analysis**

The overview HAADF-STEM image of the 100 nm thick $VO_2$ film on sapphire is shown in Fig. S1(A). High-resolution HAADF-STEM images of $VO_2$ film in the bulk and on the surface is shown in Fig S1 (B and C) respectively, with corresponding FFTs in the inset. The film is oriented along [010] (Zone axis: [100]), and there are no resolvable lattice parameter changes from the bulk of the film to the surface.
Spatially resolved Electron Energy Loss Spectroscopy (EELS) was employed to examine the oxidation state variation of V in the $VO_2$ film. Spectrum images were recorded from the interface between sapphire and $VO_2$ film, and close to the surface V $L_{2,3}$-edges are shown in Fig. S1(F). $L_3$ and $L_2$ peaks arise at 518.5 eV, 525.1 eV at the interface, and at 517.6 and 524.2 eV at the surface. These peaks correspond to the energy loss caused due to excitation of $2p_{3/2}$ and $2p_{1/2}$ states, respectively to unoccupied 3d bands(*2*). O K edge (appears as two broad peaks in 528-550 eV) and can also be observed alongside V $L_{2,3}$. Peak position and intensity ratio of $L_3/L_2$, also known as white line ratio, is a good indicator of the oxidation state of V(*3–5*). The appearance of V $L_{2,3}$ peaks at lower values at the surface compared to the interface, indicates that V at the surface is at a lower oxidation state compared to the interface. To get more insight into the variation of the peak position and white line ratio along the film, a depth profile EELS analysis was performed. The spectrum image near the surface is summed within 0.81 nm thickness, and non-linear least square (NLLS) fitting(*6*) using DM software was performed to extract the intensity and peak position of $L_3$ and $L_2$ edges. The extracted peak positions are shown in Fig. S1(G). $L_2$ peak (for e.g.) evolves from a peak position of 525.1 eV to 524.2 eV over a length scale of ~8 to 10 nm near surface (from 12-15 nm away from the surface to ~5 nm away from the surface). $L_3$ also follows a similar trend. The first 5 nm close to the surface shows saturation in the $L_2$ (also $L_3$) peak positions at 524.2 eV. White line ratio ($I_{L3}/I_{L2}$) also increases on moving towards the surface as shown in Fig. S1(H). This is also consistent with a decrease in Vanadium oxidation state when proceeding towards the surface from the bulk (*3–5*). Similar NLLS fitting and depth profiling was performed on V $L_{2,3}$ lines from the Sapphire interface to 15 nm into the bulk of the film (Total SI of 20 nm including 5 nm of Sapphire) (Fig S1(E)). The peak positions and intensity ratio, Fig. S1 (I, J), don't change, revealing no oxidation state changes in V in the bulk of $VO_2$. <u>All in all, EELS data reveals that the film is reduced in the first 12-15 nm (5 nm of constant peak position + 10 nm of gradual change) close to the surface.</u>

That the oxidation state of Vanadium on the surface is less than that in the bulk is also consistent with a slight left shift of O-K edge from surface to the bulk (Fig. S1(F)). It must be noted that we hesitate here to give the exact oxidation states, given that the fine structure is more complex, and would require detailed theoretical support. By comparing the values of peak positions studied in literature (*4*), we believe that the V in the entire film is in an oxidation state close to 4 (~4 in bulk and 4-δ at the surface). This is further substantiated by our van der Pauw measurements, that clearly show a transition at 68°C corresponding to $VO_2$ (or V in 4+), and at 61°C, corresponding to V in 4-δ. This allows us to model our thin film as two parallel layers, $VO_{2(1-\delta)}$ with a thickness of 10 nm (Fig. 1(C)), and $VO_2$ with a thickness of 90 nm, electrically in parallel to each other. Such visualization allows us to model the multiple NDR regions that we observe experimentally.

**S2. Electro-thermal Modelling (ETM)**

We modelled the electrical circuit with two variable resistors, $R_{d1}$ and $R_{d2}$, corresponding to the two layers discussed above (Fig. 1C). The thermal circuit consists of two input heat sources, $Q_1$ and $Q_2$ corresponding to Joule heating in the two layers.

ETM simulations were adapted from the single layer system in (*6*) to the two-layer system. The transition temperatures of these two variable resistors were kept as 335 K and 340 K (As confirmed from Fig. S2).

In the simulated electrical circuit illustrated in Fig. S3(A), current ($I_d$) is applied to the device with one fixed resistor ($R_x$), two variable resistors ($R_{d1}$ and $R_{d2}$) and a capacitor ($C_d$, device + parasitic). The thermal circuit in Fig. S3(B) is similar to (*7*) other than the additional heat source with its corresponding thermal resistors and capacitors for the second layer. The output power of electrical circuit in Fig. S3(A) ($Q_1(I_{d1}(R_{d1}) * V_d)$) is fed to the node A of thermal circuit while the output power $Q_2(I_{d2}(R_{d2}) * V_d)$ is fed to the node B of the thermal circuit.

Subscript 1 in the notation for thermal resistors and thermal capacitors indicates the region of the device(channel) under the active area while subscript 2 indicates the region outside the channel. Here, RV and CV refer to the thermal resistors and capacitors of reduced $VO_2$ respectively while RVO and CVO refer to the thermal resistors and capacitors of bulk $VO_2$ respectively. RS and CS respectively refer to the thermal resistors and capacitors of Sapphire (or $TiO_2$) substrate. To maintain the symmetry in the circuit, capacitors are connected in the middle of two thermal resistors as per standard practice. In the thermal circuit, the ambient temperature of the sample was simulated using a DC voltage source. In particular, the illustrated model has a 300 V DC voltage source representing ambient temperature of 300 K. The two NDRs in the DC I-V characteristic obtained using this electro-thermal model are shown in Fig. S3(C). It should be noted here that no artificial fitting parameters were used to arrive at this characteristic and that the I-V charateristic emerges naturally from the model by utilising physically measured properties. The specific details and values for each parameter used in the model is given in Table S1.

## S3. Simulated current oscillations

The ETM model leads to tonic spiking behavior as shown in Fig. S3(D). Using our electro-thermal modelling approach, we also simulate tonic burst-like behavior as seen in our measurements. The electrical circuit used to achieve this behavior is shown in Fig. S6(A) has L||C which is used to simulate the floating terminal of the cables in the measurements. The thermal circuit and the values of components used are same as in Fig. S3(B). The values for the components used in the electrical circuit have been listed in Table S2. The current output depicting tonic burst like behavior is simulated for a given voltage pulse in Fig. S6(B).

## S4. Synchronized setup for Current and Displacement Measurements

To study the coupled effect of oscillations, we measured the out of plane displacement of $VO_2$ using laser interferometry on a laser doppler vibrometer (LDV) with 632 nm laser. To simultaneously measure oscillations, both PA and LDV were synchronized (measurement set-up shown in Fig. S7). Double beam laser interferometry is a widely used technique for measuring mechanical motion detected as changes in the optical path difference at the photodetector. However, on transparent samples that show refractive index modulations, refractive index modulations also contribute to the optical path difference modulations. In our case, $VO_2$ is transparent to light in both insulating and metallic phases of $\lambda_{incident}$= 632 nm ($\lambda_{plasma}$= 909 nm in the metallic phase > $\lambda_{incident}$)(*8*) .Thus our optical path difference modulations include both refractive index modulations and mechanical oscillations in $VO_2$ across the phase transition.

The raw time domain signal of effective displacement shows a significantly high noise floor (~100 pm). In order to extract the signal more effectively, we employed a time domain filtering technique using the 'designfilt' and 'filtfilt' functions in MATLAB. The 'designfilt' function is used to design the filter by specifying its characteristics and returns a filter object that represent the designed filter. Whereas 'filtfilt' is used to apply filter to the input signal. The objective was twofold: to effectively suppress the noise components while preserving the integrity of the harmonic responses, and to extract the output signal representing the effective displacement by summing over the harmonics displacement signals.

Figs. S9 (A,E and I) shows the as obtained experimental effective displacement response with their normalized FFT's in Figs. S9 (B,F and G). Figs. S9 (C, G and K) shows the comparision between the original and the filtered effective displacement response.
Furthermore, we performed FFT on filtered signal and visualized the results in Figs. S9 (D, H and L). These plots demonstrates the time domain filtering on the frequency content, revealing the improvements in the signal to noise ratio post filtering.

Refractive index modulations in effective displacement:
These modulations were modeled in a simple transfer matrix formulation described in the manuscript. From Fig. 4E, $E_{r1}$ gives contribution to only mechanical motion whereas $E_{r1}+E_{r2}+\ldots\infty$ terms contributes to both optical and mechanical oscillations. We used the values of N and K as given by Eqn's. S1 and S2:

$$N_{VO_2} = 2.325 + 0.115 \{\sin(wt) + 0.75(sin2wt) + 0.5(sin3wt) + 0.4(sin4wt)\} \quad \text{(Eq. S1)}$$
$$K_{VO_2} = 0.244 - 0.036 \{\sin(wt) + 0.75(sin2wt) + 0.5(sin3wt) + 0.4(sin4wt)\} \quad \text{(Eq. S2)}$$

The values of $n_1$ and $n_2$ in Eqs.1 and 2 are used from(*9*) and fourier components from Fig. 2D.

To estimate the effective path difference from the electric field component ($E_\infty$, interference of all the electric field components), we separate real and imaginary part of $E_\infty$. Phase difference between $E_0$ and $E_\infty$ is given by Eq.S3, Velocity and Displacement (Eq. S4 (A) and (B)) are estimated from phase difference as in (*10*) :

$$Phase\ difference = \tan^{-1} \frac{imag(E_\infty)}{real\ (E_\infty)} \quad \text{(Eq. S3)}$$

$$Velocity = \frac{\frac{d}{dt}(Phase\ difference)*\lambda}{2*\pi} \quad \text{(Eq. S4(A))}$$

$$Displacement = \int (Velocity)\ dt \quad \text{(Eq. S4(B))}$$

Here, λ= wavelength of the incident light. Path difference is the effective displacement which is measured by the LDV.

**S5. Estimation of $d_{13}$* from Voltage oscillations**

To estimate the effective $d_{13}$* coefficient, we measured voltage and current oscillations via oscilloscope (Fig. S12(A)) with constant bias current on device with channel length 278 nm. Here, constant current (I, corresponding to NDR region) was applied to the device under test (DUT), voltage across the DUT (with 1MΩ termination) and current flowing through DUT (with 50 Ω termination) were measured (Fig. S12 (B)). It must be noted that the amplitude of current oscillations (~7mA) is much higher than the input current (0.25 mA). The presence of parasitic and device capacitances contributes to the observed phenomenon as explained next in S6.
We measured voltage oscillations for 35 cycles (1 cycle of 5ms) shown in yellow region of Fig. S12 (D). Following this, we performed DC-IV cycling for 700 cycles as shown in white region of Fig. S12(C), during which threshold voltage for insulator to metal transition fluctuated between 1.3 and 1.4 V, and metal to insulator transition fluctuated in a very narrow range between 0.35 and 0.38 V. Voltage amplitude fluctuations measured before (35 cycles, Fig. S12(D)-yellow) and after (36[th] and 37[th] cycle, Fig. S12 (D), green) DC IV cycling are less than 3%. From this data, we estimated $d_{13}$* 660 $\pm$20 pm/V.

## S6. Explanation for the amplified current across load resistor

The circuit used to explain this phenomenon is shown in Fig. S13(A) along with the current controlled I-V in Fig. S13 (B). The capacitors $C_1$ and $C_3$ are considered as parasitic capacitors while $C_2$ represents device capacitance. These are inherent to the system and cannot be eliminated during the analysis.

Initially, when the biasing current (corresponding to NDR region) is applied to the circuit, the DUT is in high resistance state. Capacitors $C_1$ and $C_2$ begin to charge, and voltage across DUT starts increasing from $V_h$ (hold voltage) to $V_{th}$ (threshold voltage).

As the voltage reaches Vth, the DUT undergoes a sudden transition from an insulating state to a metallic state. This transition results in huge current flowing through the DUT and subsequently through the load resistor, $R_l$ and the capacitor $C_3$ gets charged as well. The current is supplied by the discharge of capacitor $C_1$ and $C_2$.

At this stage, all the capacitors including $C_1$ start getting discharged through resistors, causing the voltage across DUT to drop from $V_{th}$ to $V_h$. Simultaneously, with much lower heat being generated by the device due to low resistance state, the temperature starts dropping.

At $V_h$, the DUT switches back to its insulating state due to temperature dropping below the transition temperature, causing the current passing through the load resistance to diminish. From here on out, the cycle repeats.

**Supplementary Figures**

**Fig. S1**

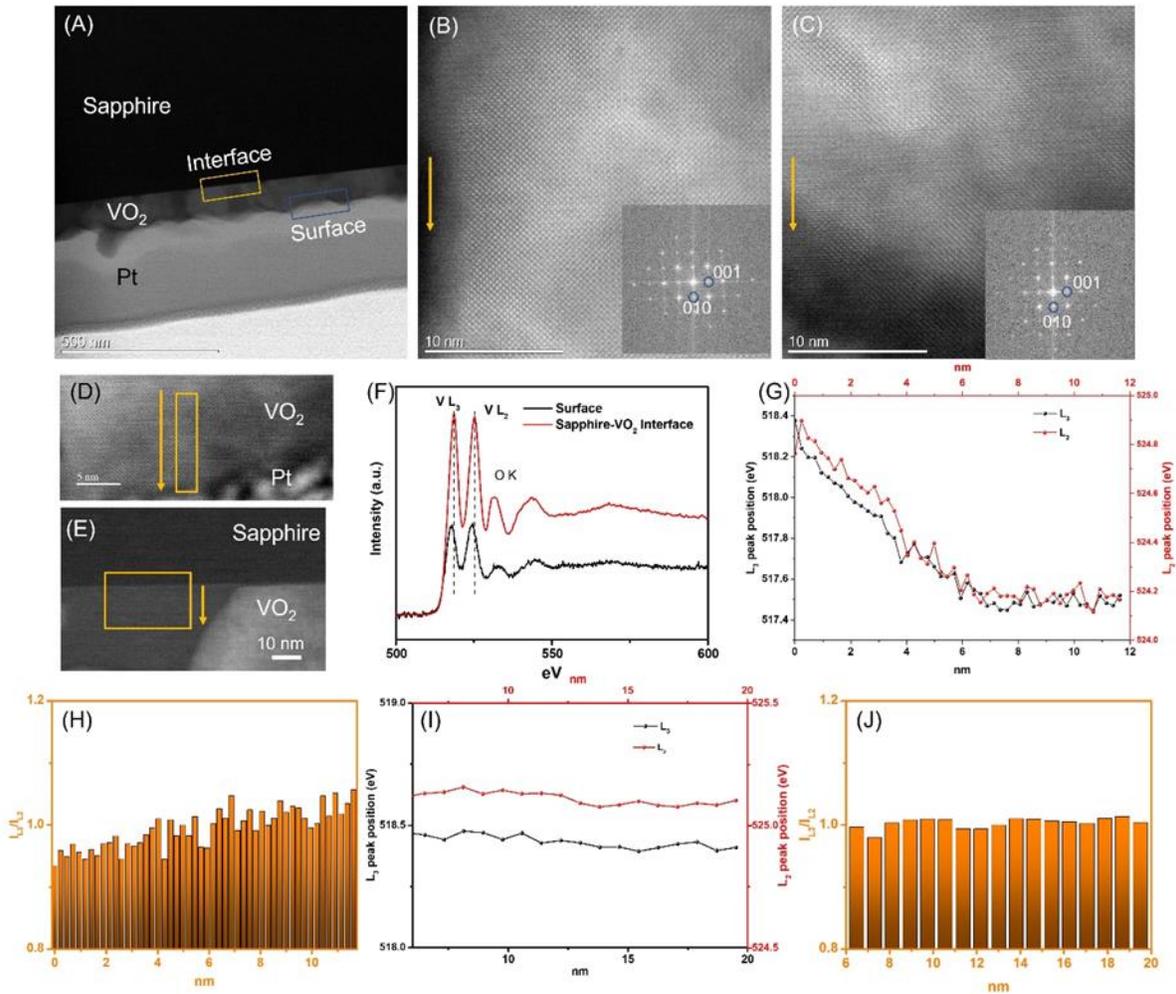

**STEM-EELS analysis**: (A) Low-magnification HAADF-STEM image, different regions of FIB lamella are marked (B) High-resolution HAADF-STEM image within the thin film (Inset: FFT. (C) High-resolution HAADF-STEM image from the surface (Inset: FFT), Marked arrows show the growth direction of the film. (D) EELS spectrum imaging (SI) area on the surface (E) EELS spectrum imaging (SI) area near the interface. (F) Summed Spectra of the interface and bulk surface. (Depth profile analysis direction is marked by arrow) (G) $L_2$ and $L_3$ peak positions extracted using NLLS fitting of the bulk surface SI. Depth profiling follows the direction of the arrow shown in (D). (H) White line ratio and peak separation between $L_2$ and $L_3$ edge of V on the bulk surface. (I) $L_2$ and $L_3$ peak positions extracted using NLLS fitting of the interface SI. (J) White line ratio and peak separation between $L_2$ and $L_3$ edge of V on the interface.

**Fig. S2**

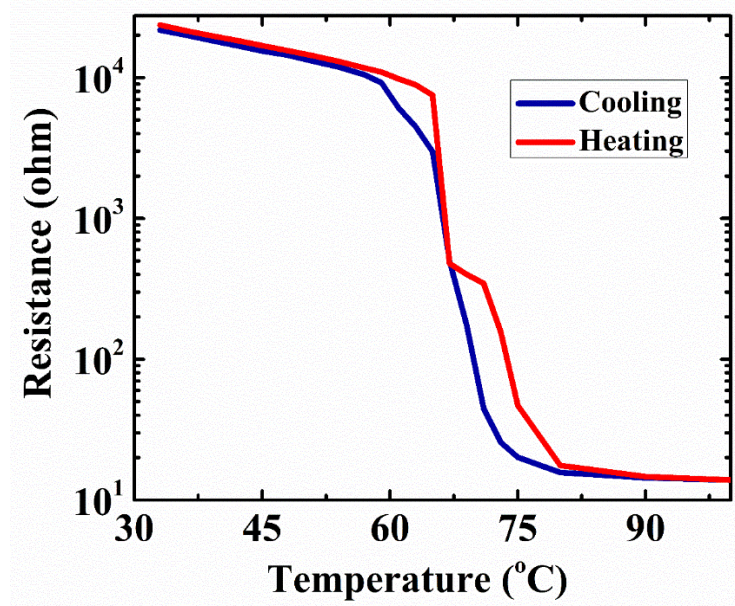

**Resistance vs temperature of a lateral two-terminal device:** This shows double hysteresis loops indicating the presence of two transition temperatures. Transition temperatures extracted as the mid points of these sub-loops are about 61 and 68 °C respectively. 68 °C is also bulk $VO_2$ transition, which in conjunction with EELS data suggests that the majority of the film is $VO_2$, and at the surface is reduced to an oxidation state(s) that lowers the transition temperature very slightly.

**Fig. S3**

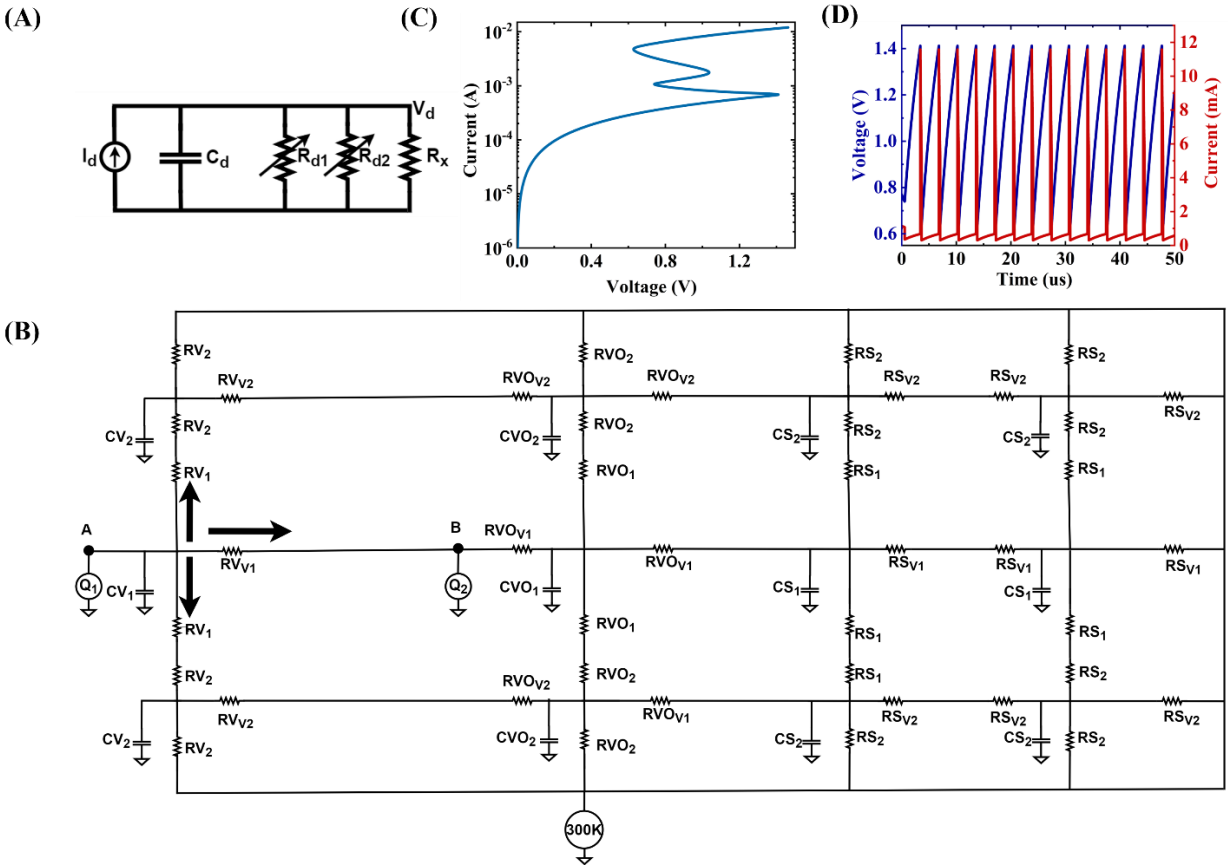

**Electro-thermal simulations.** (**A**) Electrical circuit with two variable resistors ($R_{d1}$ and $R_{d2}$). The two variable resistors are used for two different oxidation states of Vanadium. $R_x$ is the fixed core-shell resistance (resistance outside the filament). $C_d$ is the combined device and parasitic capacitance. Current ($I_d$) is applied to the circuit and voltage across the channel is measured as $V_d$. (**B**) Thermal circuit with two heat sources $Q_1$ and $Q_2$, one for reduced $VO_2$ and other for $VO_2$. The ambient temperature (ground potential) was set at room temperature (300 K). (**C**) Simulated DC I-V characteristic obtained from pumping current and measuring voltage ($V_d$) using the circuit in (A). (**D**) Simulated voltage and current periodic (Tonic spiking) oscillations with input as constant current bias of 1mA.

Fig. S4

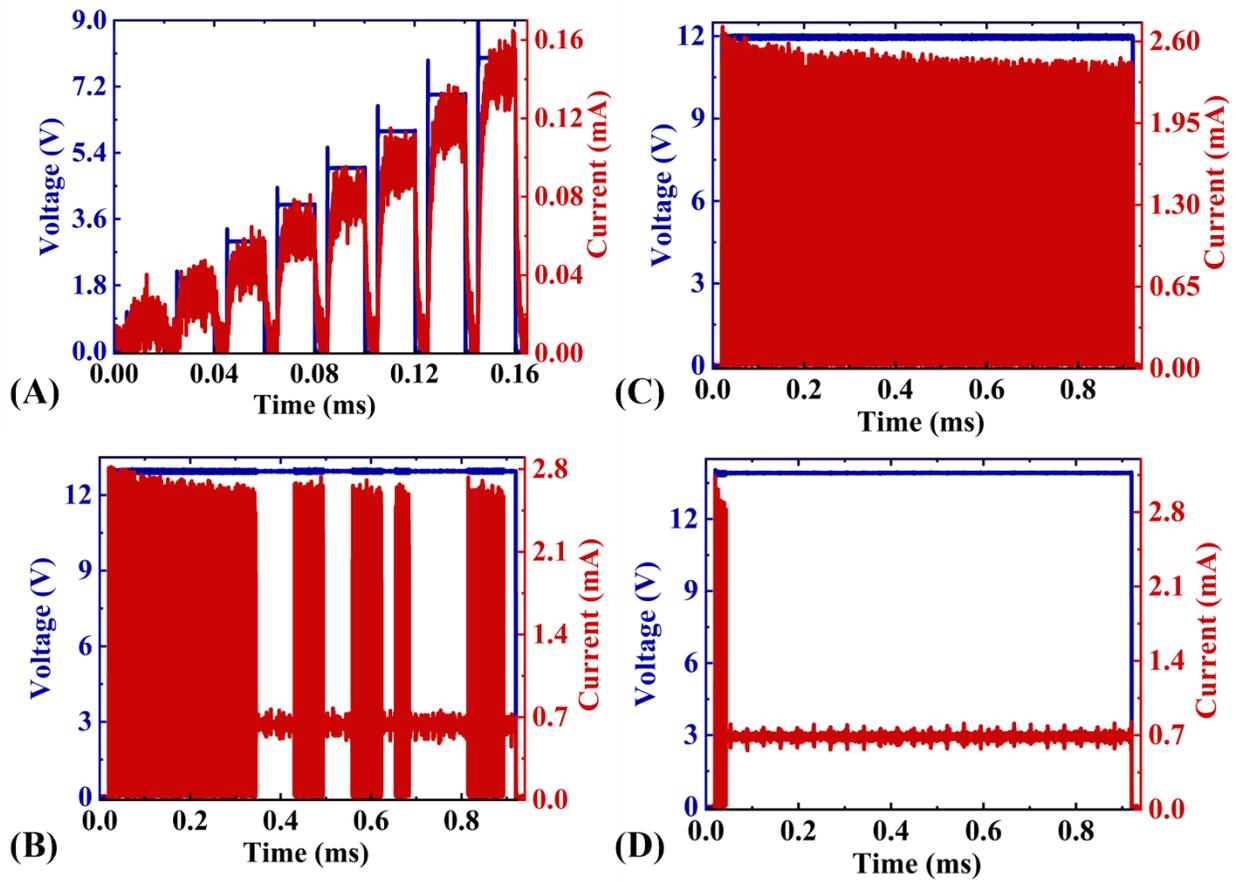

**Measured different types of current oscillations. Y-axis left- Voltage applied. Y-axis right-Current measured.** (**A**) Current at input voltage less than 9V. (**B**) Current oscillations (Tonic spiking) at 12 V. (**C**) Current oscillations at 13.2 V (Tonic Bursting). (**D**) Current response at 14V (Phasic spiking).

**Fig. S5**

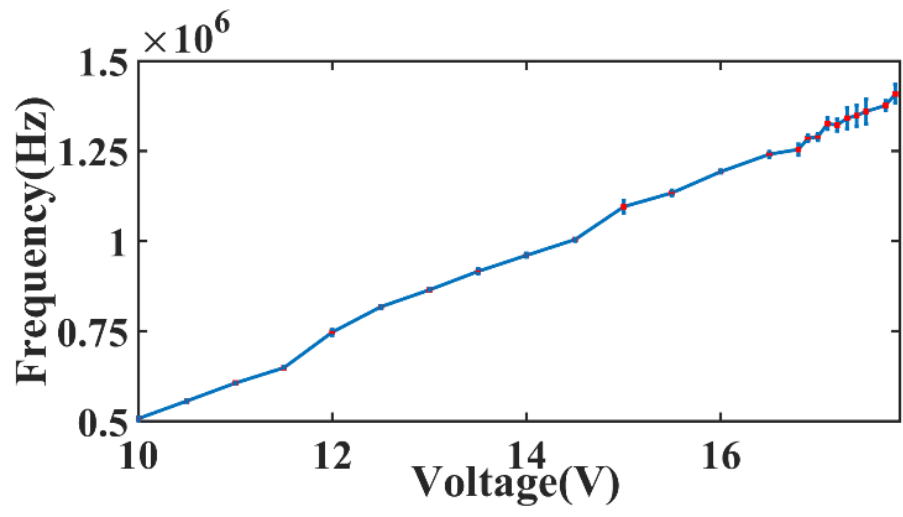

Frequency of current oscillations at different voltages.

**Fig. S6**

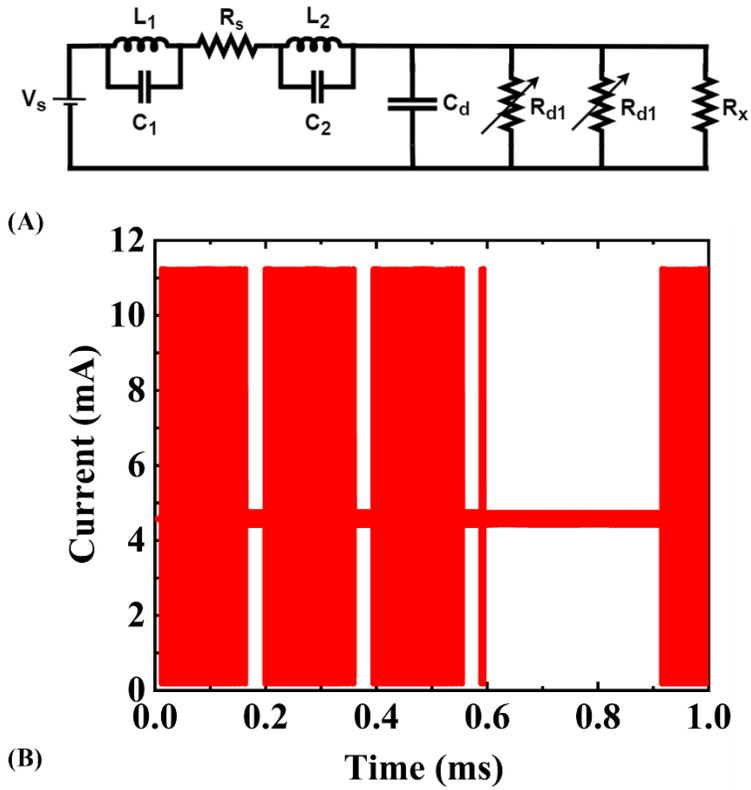

**Simulated Tonic-burst like behavior.** (**A**) Electrical circuit used for simulating tonic bursting current oscillations. L and C are the inductors and capacitors which act as a floating terminal of the cables. (**B**) Simulated Current response from the circuit in (A) showing tonic-burst like behavior.

**Fig. S7**

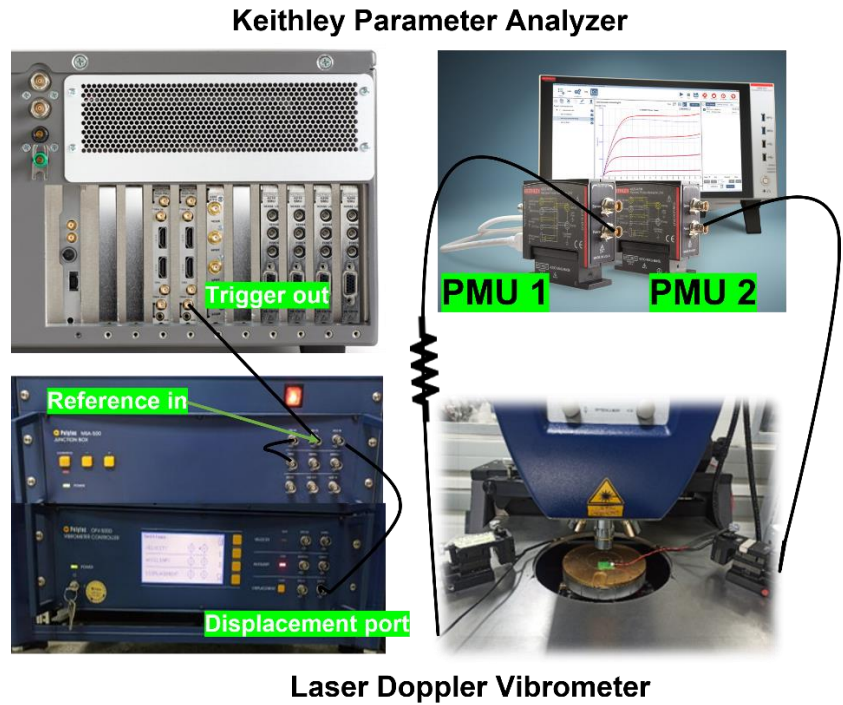

**Synchronized LDV and Keithley parameter analyzer setup.** Both the instruments were synchronized by connecting trigger out channel of PA to the reference in port of LDV.

**Fig. S8**

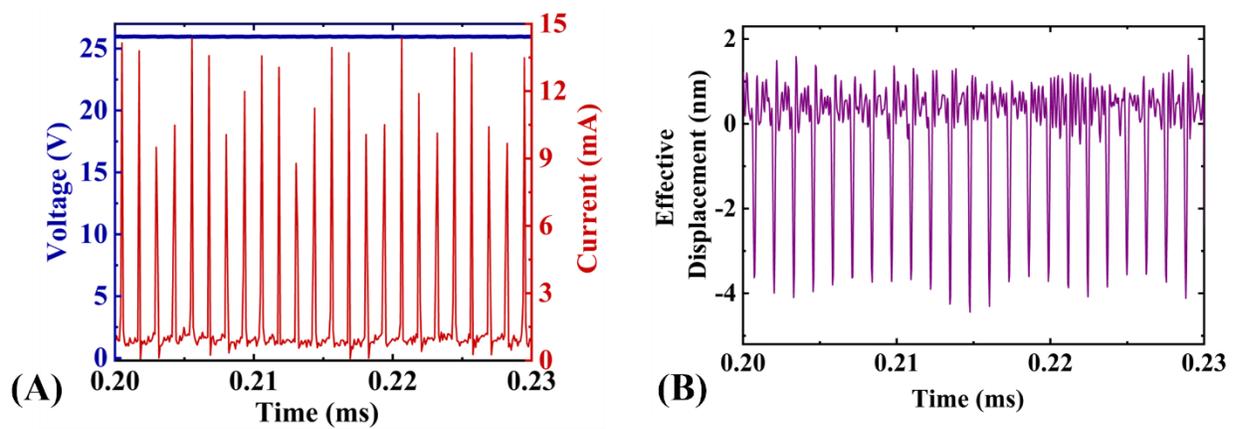

**Zoomed-in images of** (**A**) Current oscillations in Fig. 2(B). (**B**) Effective displacement oscillations in Fig. 2(C).

**Fig. S9**

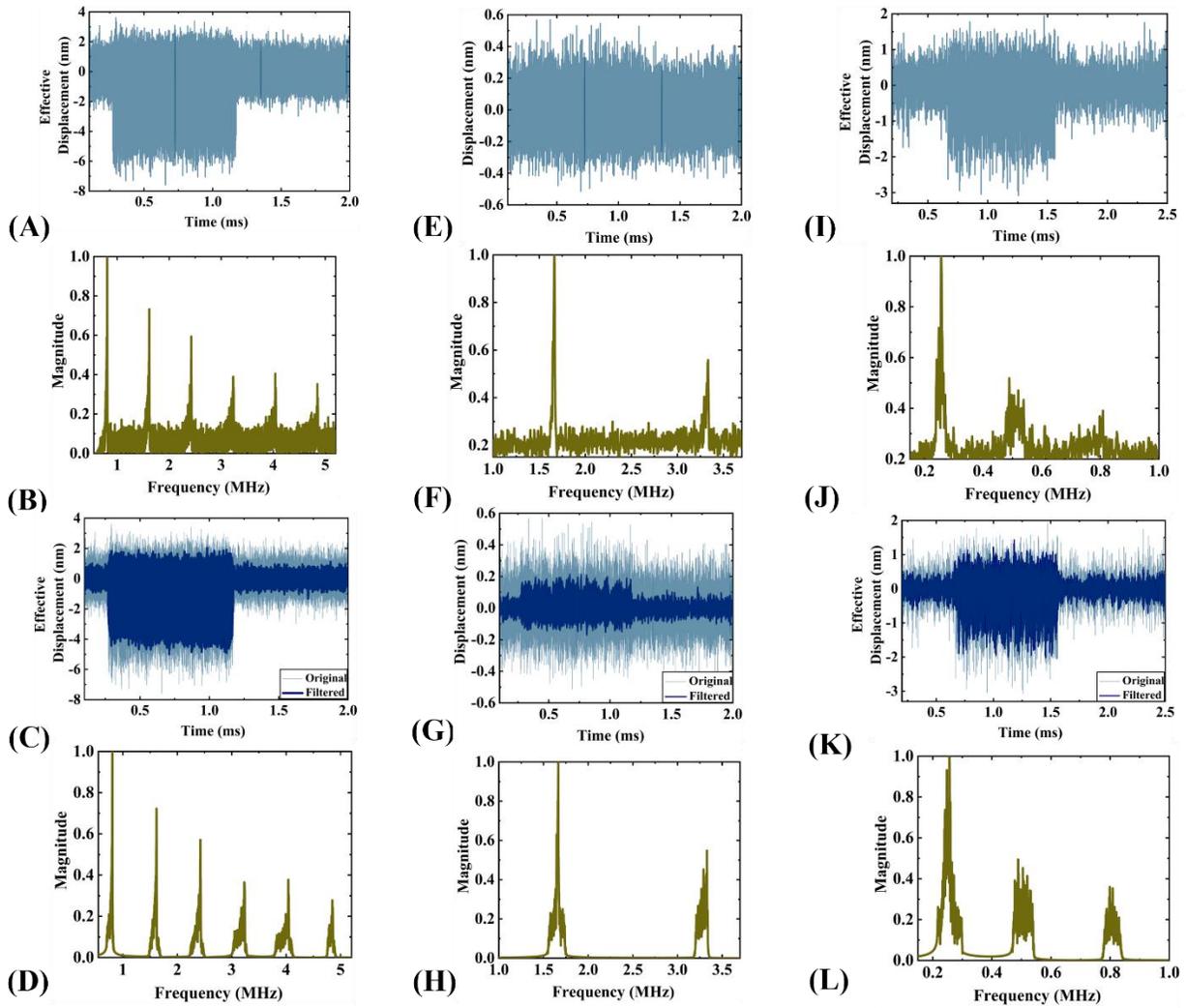

**Comparison of original and filtered effective displacement response** (**A**) Effective displacement of as obtained response of 100 nm thick VO$_2$ on Sapphire. (**B**) Normalized FFT of original data of 100 nm VO$_2$ on Sapphire. (**C**) Comparison of effective displacement of as original and filtered response of 100 nm VO$_2$ on Sapphire. (**D**) Normalized FFT of filtered response of 100 nm VO$_2$ on Sapphire. (**E**) Displacement of as obtained data of 100 nm VO$_2$ on Sapphire coated with resist and Pt on top of the channel. (**F**) Normalized FFT of original data of 100 nm VO$_2$ on Sapphire coated with resist and Pt on top of the channel. (**G**) Comparison of displacement of original and filtered response of 100 nm VO$_2$ on Sapphire coated with resist and Pt on top of the channel. (**H**) Normalized FFT of filtered response of 100 nm VO$_2$ on Sapphire coated with resist and Pt on top of the channel. (**I**) Effective displacement of as obtained response of 10 nm VO$_2$ on TiO$_2$. (**J**) FFT of original data of 10 nm VO$_2$ on TiO$_2$. (**K**) Comparison of effective displacement of original and filtered response of 10 nm VO$_2$ on TiO$_2$. (**L**) Normalized FFT of filtered response of 10 nm VO$_2$ on TiO$_2$.

**Fig. S10**

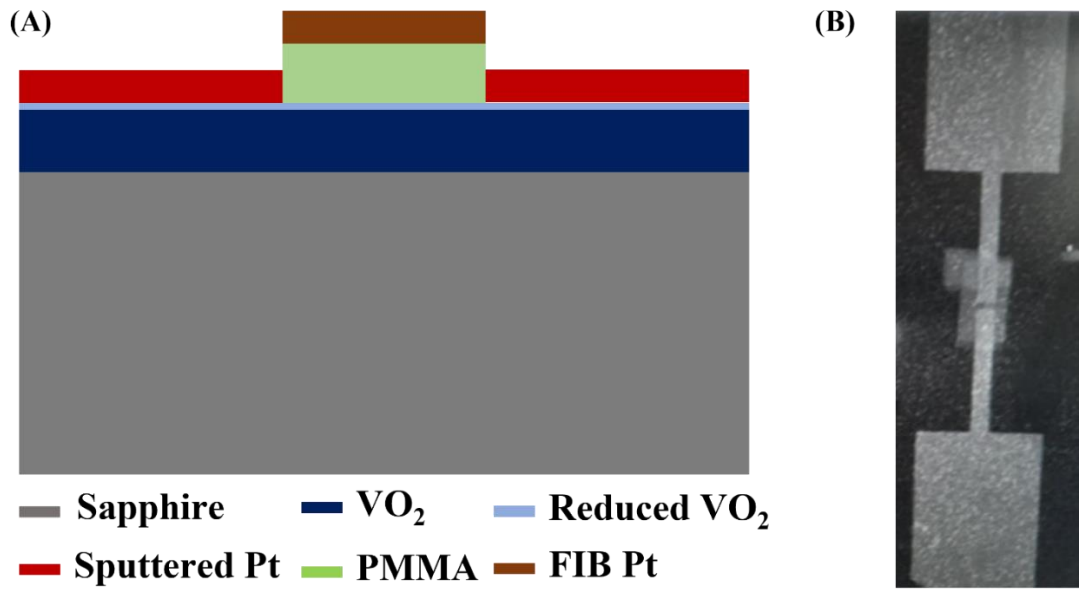

**Device fabrication to measure the response of only mechanical oscillations.** (**A**) Schematic of fabricated device. (**B**) SEM of the fabricated device.

**Fig. S11**

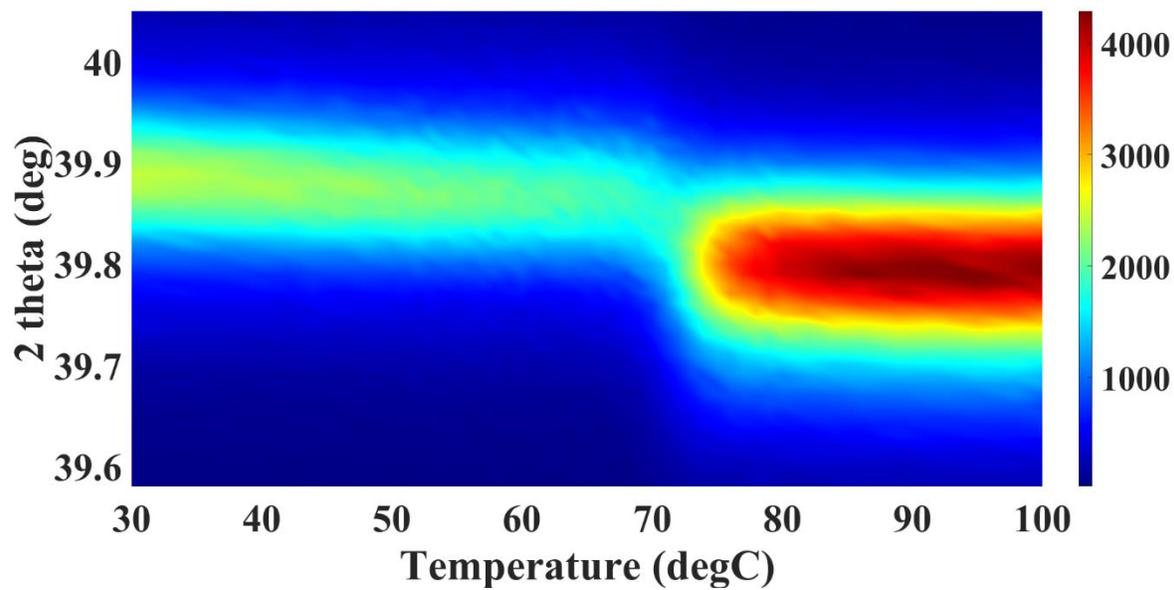

Two-dimensional temperature-dependent XRD patterns of θ-2θ. Here, 2θ is plotted as a function of temperature during heating from 30ºC to 100ºC. There is a clear shift in 2θ from monoclinic to rutile phase of $VO_2$.

**Fig. S12**

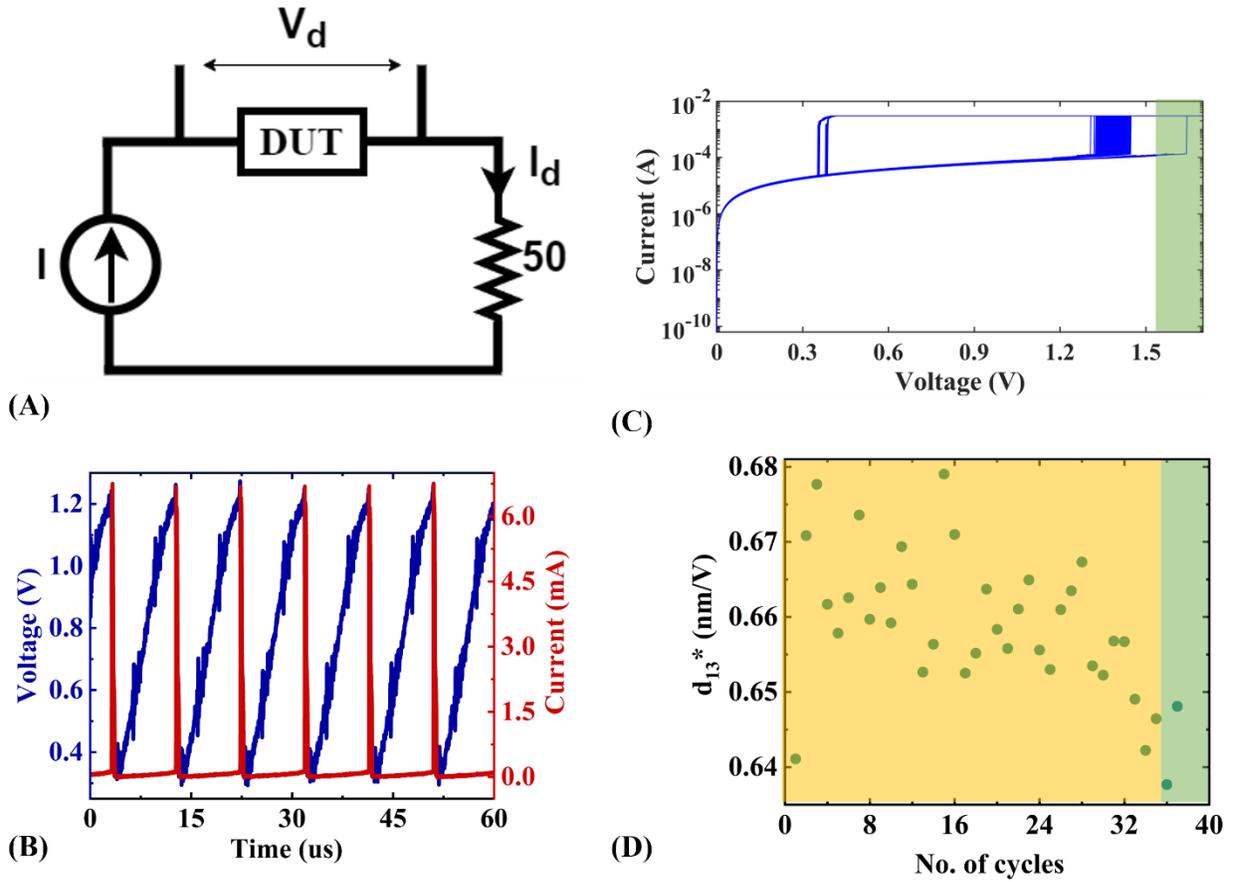

**Effective $d_{13}$ for 100 nm thick $VO_2$ film with 278 nm channel length measured via oscilloscope with constant input current source**. Voltage oscillations measured across device using differential measurement in oscilloscope (high termination resistance of 1 MΩ resistor was chosen for both ports of oscilloscope). Current oscillations measured across load resistor (50Ω). (**A**) Electrical circuit with input constant current to devices with different channel lengths. (**B**) Voltage and current oscillations with constant input current of 250uA. (**C**) DC-IV characteristics (V-controlled) with 700 cycles. The cycle marked with green is after all acquiring all the voltage oscillations runs. (**D**) Effective $d_{13}$ of initial 35 cycles (marked with yellow) and 36,37 cycles after 700 DC-IV cycles (marked with green).

**Fig. S13**

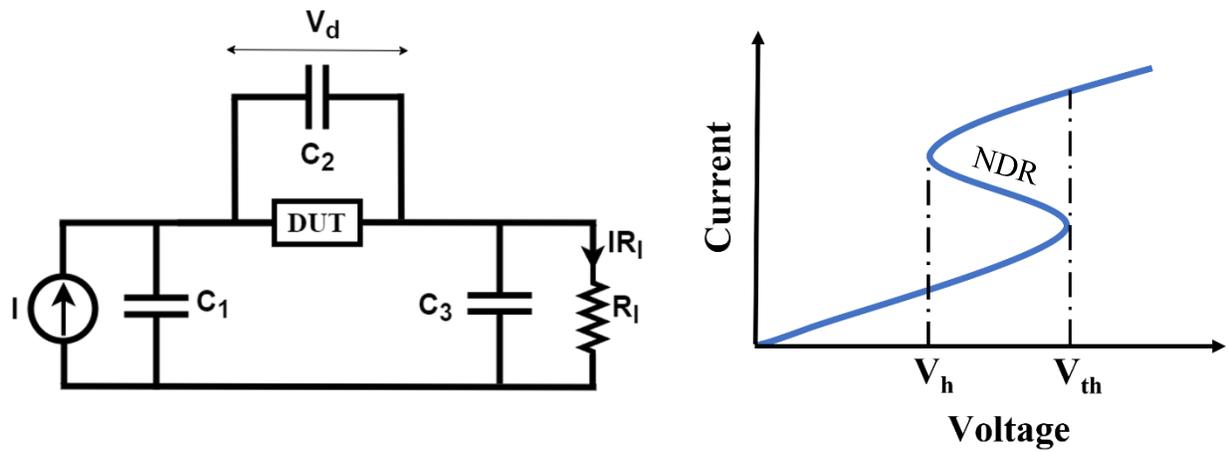

**Circuit used to explain the amplification of current flowing across load resistor ($R_l$) (A)** The circuit configuration involves the application of constant current to the circuit, which includes DUT and $R_l$ (50Ω). The load resistor is used to measure voltage drop and hence to output current. The values of $C_1$-$C_3$ determine the frequency of oscillation. **(B)** The I-V characteristic of DUT with NDR region between $V_h$ and $V_{th}$.

**Fig. S14**

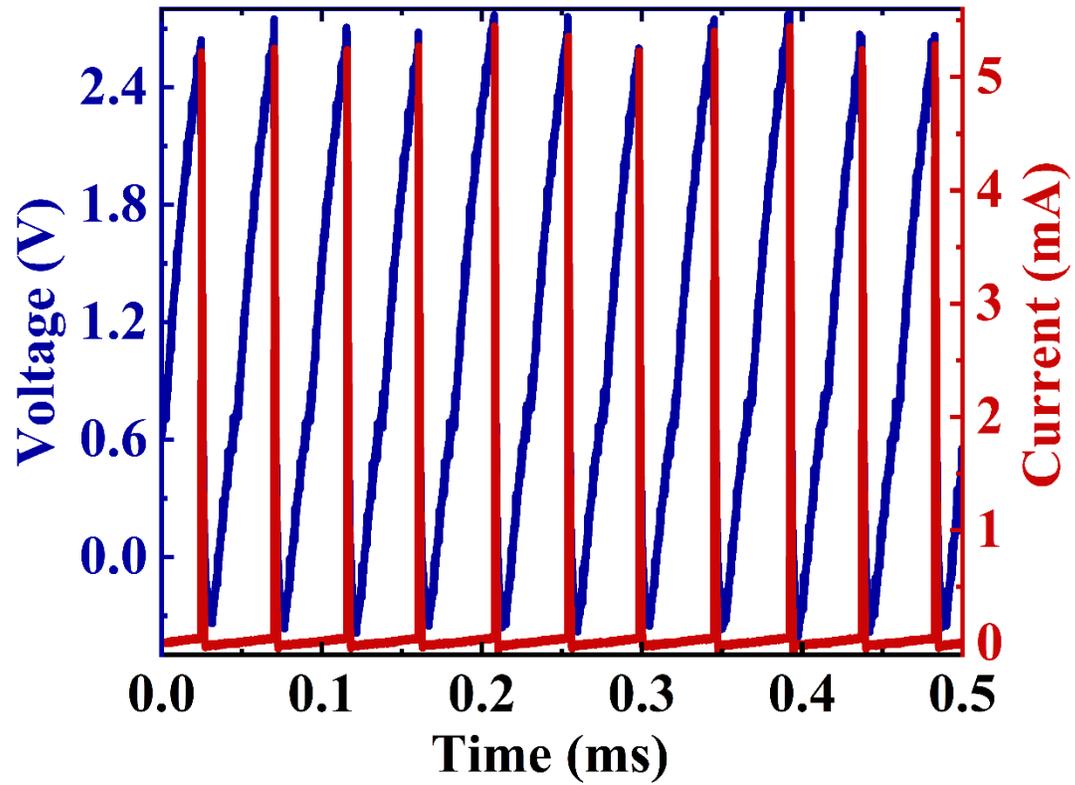

**Voltage and current oscillations of device with 600 nm channel length (VO$_2$ (10 nm) on TiO$_2$).** For ΔV-max =3.11 V and Δn = 0.115, Electric field (E-max) = 5.18 MV/m and r$_{13}$= Δn/E-max= 22 nm/V.

**Fig. S15**

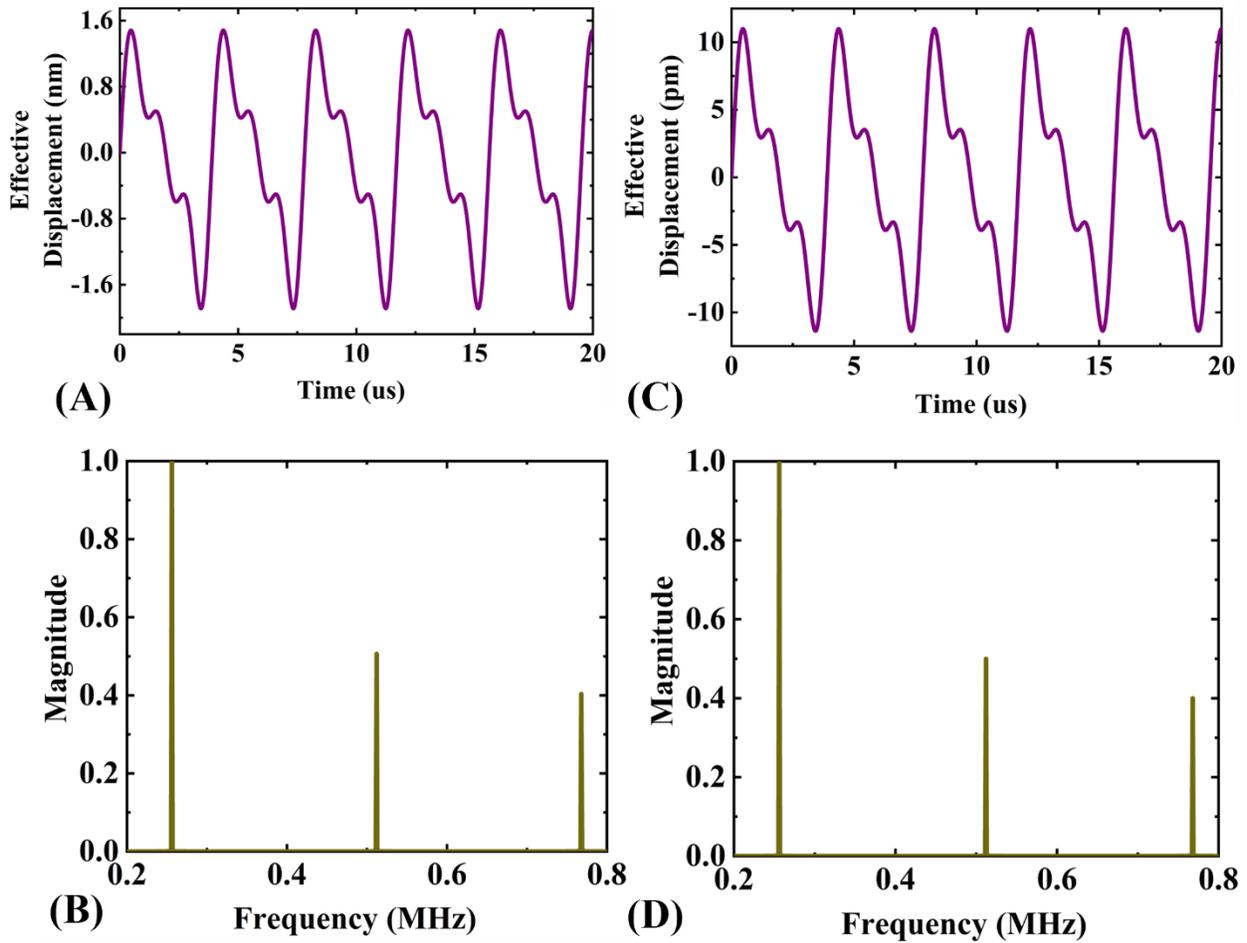

**Analytical simulations** (**A**) Effective displacement considering infinite reflections of VO$_2$ (10 nm) on TiO$_2$. (**B**) Normalized FFT corresponding to (A). (**C**) Effective displacement considering only E1 of VO$_2$ (10 nm) on TiO$_2$. (**D**) Normalized FFT corresponding to (C).

**Table S1**

| Parameters | Description | Value | Reference |
|---|---|---|---|
| $\sigma_{ins-V}$ | Insulating state electrical conductivity of reduced vanadium oxide | $1e^5$ S/m | Estimated |
| $\sigma_{m-V}$ | Metallic state electrical conductivity of reduced vanadium oxide | $5e^5$ S/m | Estimated |
| $T_{s-V}$ | Transition temperature of reduced vanadium oxide | 335 K | Estimated |
| $\alpha\_V$ | Spread about transition temperature of reduced vanadium oxide | 1 | Estimated |
| $\kappa_{ins}$ | Insulating state thermal conductivity | 3.6 W/m/K | (*11*) |
| $\kappa_m$ | Metallic state thermal conductivity | 5.4 W/m/K | (*11*) |
| $\rho$ | Mass density of both vanadium layers | 4571 kg/m$^3$ | |
| C | Heat capacity of both vanadium layers | 690 J/kg/K | (*12*) |
| **Thickness_V** | Thickness of reduced vanadium oxide | $10e^{-9}$ m | Estimated |
| $\sigma_{ins-VO}$ | Insulating state electrical conductivity of vanadium oxide | 1e3 S/m | Estimated |
| $\sigma_{m-VO}$ | Metallic state electrical conductivity of vanadium oxide | 3e5 S/m | Estimated |
| $T_{s-VO}$ | Transition temperature of vanadium oxide | 340 K | Estimated |
| $\alpha\_VO$ | Spread about transition temperature of vanadium oxide | 4 | Estimated |
| **Thickness_VO** | Thickness of vanadium oxide | $90e^{-9}$ m | Estimated |
| **Channel width** | Width of the device | $2e^{-6}$ m | Fabricated |
| **Filament width** | Width of filament formed | $0.1e^{-6}$ m | Fabricated |
| **Sample Length** | Length of the sample | $1e^{-2}$ m | Fabricated |
| **Sample Width** | Width of the sample | $1e^{-2}$ m | Fabricated |
| **ρ_sapphire** | Mass density of Sapphire | $3.97e^3$ kg/m$^3$ | (*13*) |
| **C_sapphire** | Heat capacity of sapphire | 763 J/kg/K | (*13*) |
| **Thickness sapphire** | Thickness of sapphire | $250e^{-6}$ m | |
| **C$_p$+C$_d$** | Parasitic and device capacitances | $2e^{-9}$ F | Estimated |

The table provides the description and the values of the parameters used in the ETM for reduced VO$_2$, VO$_2$ and sapphire substrate. Here, V is used as an abbreviation for reduced VO$_2$ layer and VO for VO$_2$.

**Table S2**

| Parameters | Value |
|---|---|
| $V_s$ | 47.64 V |
| $R_s$ | 10 kΩ |
| $L_1$ | 0.1 H |
| $L_2$ | 0.01 H |
| $C_1$ | $6e^{-9}$ F |
| $C_2$ | $1e^{-5}$ F |
| $C_d$ | $1.5e^{-9}$ F |

13. S. Al, SAPPHIRE ($Al_2O_3$), 213.